# High-quality intrinsic circular dichroism in silicon-based bound states in the continuum metasurfaces via strong coupling*

ZHU Leyi[1, 2], SANG Tian[1, 2], YANG Guanhua[1, 2], WANG Chenyang[1, 2]

1. School of Optoelectronic Information and Physical Science, Jiangnan University, Wuxi 214122, China
2. Jiangsu Provincial Research Center of Light Industrial Optoelectronic Engineering and Technology, Wuxi 214122, China

**Abstract** Bound states in the continuum (BICs) can significantly enhance spin-selective light-matter interactions. However, current BIC metasurfaces generally rely on single or multiple chiral BIC modes to enhance circular dichroism (CD), limiting their application in functional integrated devices. This paper proposes a method that utilizes the strong coupling effect in BIC metasurfaces to achieve intrinsic CD with both a high quality factor ($Q$-factor) and high efficiency. The metasurface is designed as a silicon-based cross-shaped nanohole array, which can simultaneously support a TM-like BIC mode and a TE-like guided-mode resonance (GMR) mode. By introducing under-etching to break the out-of-plane symmetry and tuning the in-plane asymmetry parameter, strong coupling between the quasi-BIC (QBIC) mode and the GMR mode can be induced, thereby achieving near-perfect intrinsic CD with a high $Q$-factor. In the strong-coupling region, the resonant wavelengths of the QBIC and GMR modes exhibit anti-crossing behavior, while the imaginary parts of the eigenfrequency cross; the corresponding Rabi splitting energy reaches a minimum of 3.16 meV, and the induced intrinsic CD achieves a maximum of 0.995 at a wavelength of 1653.9 nm, with a high $Q$-factor of 1083. The evolution of far-field polarization singularities shows that, under strong coupling, the $\Gamma$ point corresponds to a left-handed circular polarization singularity ($C$-point). Furthermore, the strong-coupling-induced CD exhibits good robustness. Even under considerable variations in structural parameters, the high-$Q$ and high-efficiency CD response is maintained. This method offers a novel design strategy for integrated chiral photonic devices and holds promise for applications in areas such as chiral sensing, chiral lasers, and nonlinear chiral optics.

---

**Keywords** intrinsic chirality; bound states in the continuum; circular dichroism; strong coupling

# 1 Introduction

Optical chirality refers to the differential response of nanostructures to left-handed circular polarization (LCP) and right-handed circular polarization (RCP) light[1,2]. As a core physical mechanism in modern nanophotonics, its response strength is typically quantified using circular dichroism (CD)[3-5]. However, the chiral optical responses of natural materials and conventional nanostructures are generally weak. Due to the detuning between structural dimensions and excitation wavelengths, their CD signals remain limited[6,7]. Although strong CD can be achieved in planar systems by breaking in-plane symmetry, such as by introducing structural anisotropy or employing oblique incidence, this chirality arises from specific illumination and detection geometries. This phenomenon is classified as extrinsic chirality (pseudo-chirality)[8-10] rather than an intrinsic property of the structure. Consequently, its practical applications in chiral emission, chiral sensing, and polarization photodetection are significantly restricted.

To achieve intrinsic chirality (true chirality) independent of excitation conditions, early efforts primarily relied on complex three-dimensional chiral metamaterials, such as helical structures[11-13], "L"-shaped stereoscopic units[14], or spherical cap structures[15]. While these structures can generate strong chiral responses, their fabrication processes are cumbersome. They are difficult to integrate with standard planar processes, hindering large-scale applications. In recent years, bound states in the continuum (BIC), as ideal localized states existing within the radiation continuum, have demonstrated significant research value and application potential in nanophotonics. This is due to their theoretically infinite quality factor ($Q$-factor) and infinitely long photon lifetimes[16-20]. By introducing controlled symmetry breaking, BICs can transform into quasi-BICs (QBICs) with finite radiation loss. QBICs inherit the high-$Q$ characteristics of BICs, strongly confining optical fields at subwavelength scales, thereby significantly enhancing light-matter interactions[21-23]. Achieving strong intrinsic chiral optical responses based on QBICs has become a key research direction in nanophotonics. Current mainstream implementation schemes primarily rely on geometric designs involving symmetry breaking. Examples include metasurface units with tilted sidewalls[24-26] or metasurface structures with unequal heights that simultaneously break in-plane and out-of-plane symmetries[27-29]. These designs excite high-$Q$ chiral QBIC modes at the $\Gamma$ point in momentum space. Furthermore, recent studies indicate that mode coupling systems based on BIC metasurfaces can give rise to novel optical phenomena and applications. For instance, strong coupling between two adjacent QBIC bright modes in silicon-based dimers can achieve wavelength-stable and bandwidth-tunable electromagnetically induced transparency[30]. Similarly, coupling between toroidal dipole bright modes and two types of BIC dark modes in silicon-based nanopores can realize dual-channel high-$Q$ electromagnetically induced transparency[31]. Additionally, the strong coupling effect of resonant modes in metasurfaces can be utilized to achieve high-purity chiral laser emission[32]. However, despite the significant potential of strong coupling systems in optical field manipulation, utilizing this mechanism to achieve intrinsic CD with both high $Q$-factors and high efficiency in BIC metasurfaces remains an important direction requiring further exploration.

In this context, we propose an innovative scheme to achieve intrinsic CD with both high $Q$-factors and high efficiency by leveraging the strong coupling effect in BIC metasurfaces. The proposed metasurface employs a silicon-based cross-shaped nanopore design with $C_4$ symmetry. This design simultaneously supports TM-like BIC modes and TE-like guided-mode resonance (GMR) modes in the communication band. Undercutting is used to break the out-of-plane symmetry, while precise tuning of in-plane asymmetry parameters establishes a strong coupling system between QBIC and GMR modes. This ultimately achieves a near-perfect high-$Q$ intrinsic CD response. The results indicate that within the strong coupling region, the resonance wavelengths of the QBIC#1 and GMR modes exhibit significant anti-crossing characteristics. The imaginary parts of their eigenfrequencies cross and overlap, with a corresponding Rabi splitting energy valley of 3.16 meV. Under these conditions, the QBIC#1 mode achieves a near-perfect intrinsic chirality with a CD of 0.995 at a wavelength of 1653.9 nm, while maintaining a high $Q$-factor of 1083. The evolution characteristics of far-field polarization singularities show that the strong coupling effect drives left-handed circular polarization singularities ($C$-points) to precisely locate at the $\Gamma$ point in momentum space. Furthermore, this strong coupling-induced intrinsic CD exhibits excellent structural robustness. Even with significant deviations in core structural parameters, the CD response stably maintains high-$Q$ and high-efficiency.

# 2 Structural Design and Model

## 2.1 Metasurface Structure

Figure 1(a) illustrates the schematic of a silicon-based BIC metasurface designed to achieve high-$Q$ intrinsic CD via the strong coupling effect. The structure consists of a silicon nanopore array with a period of $P$ and a thickness of $h$, placed on a fused silica ($SiO_2$) substrate. For the communication band studied in this work, the refractive indices of Si and $SiO_2$ are set to 3.48 and 1.47, respectively. Figure 1(b) shows the schematic of a single unit cell of the achiral silicon metasurface. Without perturbation, this cross-shaped structure can be viewed as comprising five square nanopore units of identical size with side length $w$. The specific structural parameters are: period $P$ = 800 nm, thickness $h$ = 470 nm, and side length $w$ = 160 nm. Figure 1(c) depicts the schematic of a single unit cell of the chiral silicon metasurface. In the $x$-$y$ plane, the length of the upper square hole along the $x$ direction is shortened to $w$-$\delta$, while the length of the lower square hole is extended to $w$+$\delta$. This breaks the in-plane mirror and $C_4$ symmetries of the structure. The in-plane asymmetry parameter is defined as $\alpha = \delta/w$. Undercutting with a thickness of $d$ is used to break the out-of-plane mirror symmetry, with the out-of-plane asymmetry parameter defined as $\beta = d/h$. Therefore, when both $\alpha$ and $\beta$ are non-zero, the in-plane and out-of-plane mirror symmetries of the structure are simultaneously broken. All numerical simulations were performed using the commercial software COMSOL Multiphysics, which is based on the finite element method (FEM). The simulation region was confined to a single unit cell, with periodic boundary conditions applied along the $x$ and $y$ directions. Perfectly matched layers (PML) were set in the $z$ direction on both the air side above the Si nanostructure and the $SiO_2$ substrate side below to simulate non-reflective infinite extension regions.

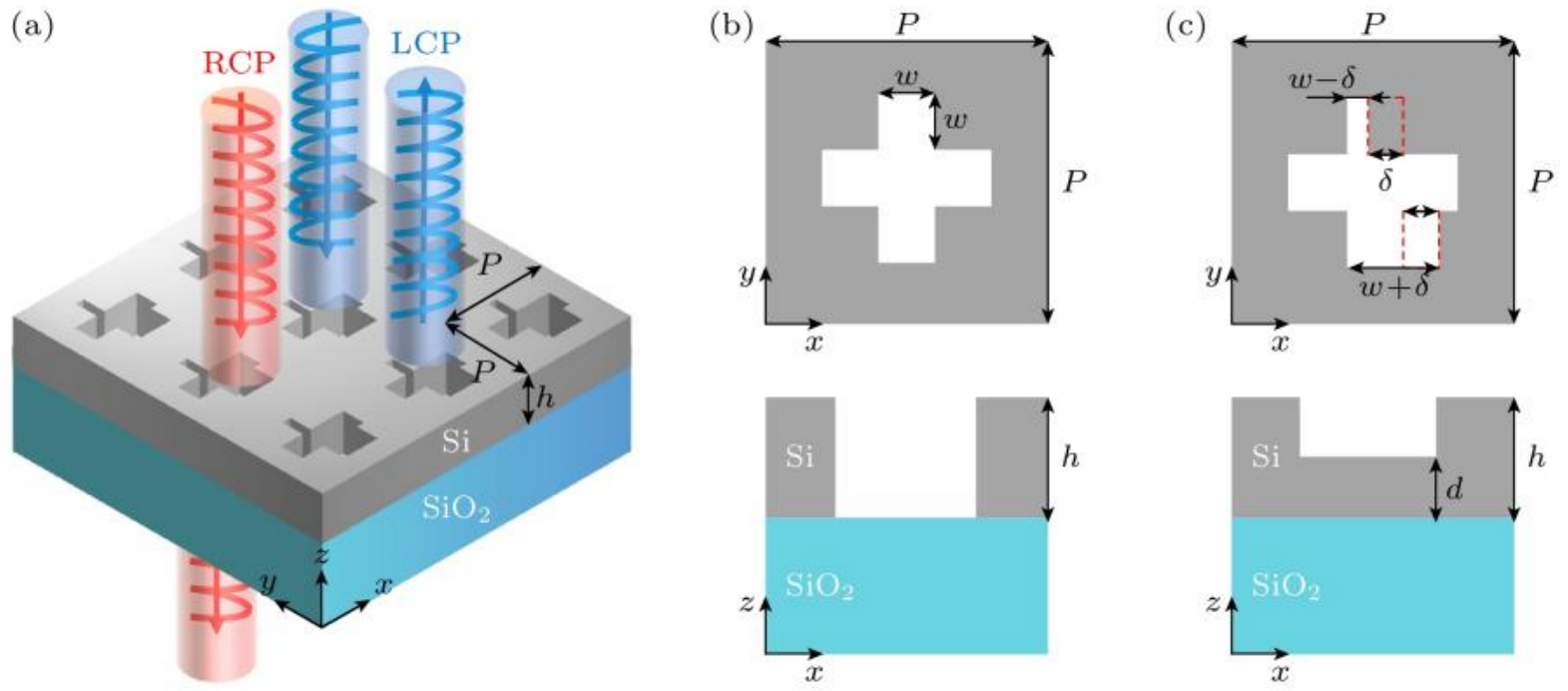

**Fig. 1 (a) Schematic of the Si metasurface; *x-y* plane (top view) and *x-z* plane (side view) of the unit cell for the (b) achiral and (c) chiral Si metasurfaces.**

## 2.2 Band Structure and Mode Characteristics

To investigate the mode characteristics of the metasurface, we first calculated the band structure, $Q$-factors, and modal field distributions without introducing perturbations, i.e., when $(\alpha, \beta) = (0, 0)$. Figure 2(a) shows the band structure of the symmetric Si metasurface. Within the frequency range studied in this work, the metasurface supports Bloch modes represented as $TM_1$, $TM_2$, and TE bands. Figure 2(b) displays the $Q$-factor distribution of these modes in momentum space. The $TM_1$ and $TM_2$ bands are associated with two symmetry-protected BIC modes at the $\Gamma$ point, where their $Q$- factors theoretically approach infinity[33-35]. As the wave vector deviates from the $\Gamma$ point, the $Q$-factors decrease rapidly. In contrast, the TE band is associated with GMR modes, exhibiting finite $Q$-factors across the entire wave vector range. Figure 2(c) illustrates the electromagnetic field distributions of different bands in the $x$-$y$ plane at the $\Gamma$ point for $(\alpha, \beta) = (0, 0)$. The $TM_1$ and $TM_2$ modes are primarily characterized by dominant in-plane magnetic fields and $E_z$ components, while the $E_x$, $E_y$, and $H_z$ components are negligible. This confirms that both are quasi-TM modes with electric field distributions exhibiting $C_4$ symmetry. The TE mode is mainly characterized by dominant in-plane electric fields and $H_z$ components, with negligible $H_x$, $H_y$, and $E_z$ components, which is typical of quasi-TE modes[36]. These results confirm that the structure can simultaneously support quasi-TM BIC and quasi-TE GMR modes.

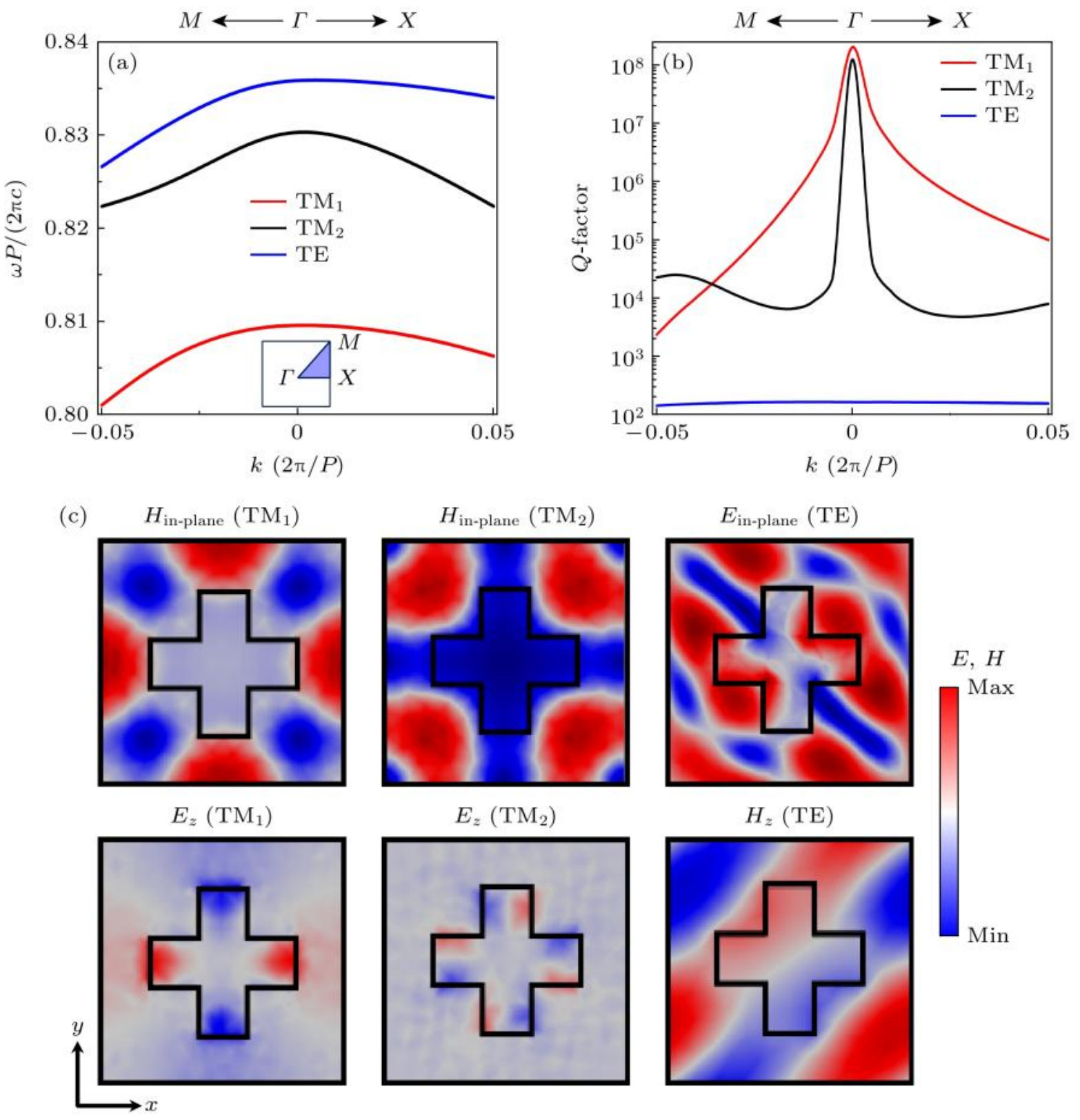

**Fig. 2 (a) Band structure, (b) $Q$-factor, and (c) mode field distribution in the $x$-$y$ plane of the Si metasurface for $(\alpha, \beta) = (0, 0)$.**

# 3 Numerical Simulation Results and Analysis

## 3.1 Analysis of Transmission CD Response

Figure 3 illustrates the optical response of the metasurface under transmission conditions, with structural parameters consistent with those in Figure 1. Here, CD is defined as the relative difference in transmittance between RCP and LCP light under

normal incidence:

$$\mathrm{CD} = \frac{T_{\mathrm{RCP}} - T_{\mathrm{LCP}}}{T_{\mathrm{RCP}} + T_{\mathrm{LCP}}}, \tag{1}$$

Here, $T$ denotes the transmittance, where $T_{\mathrm{RCP}} = T_{\mathrm{RR}} + T_{\mathrm{LR}}$ and $T_{\mathrm{LCP}} = T_{\mathrm{LL}} + T_{\mathrm{RL}}$ ($T_{ij}$ represents the output polarization $i$ converted from the input polarization $j$; $i$ = R, L; $j$ = R, L; R denotes RCP, and L denotes LCP). Figures 3(a)—(c) present the two-dimensional spectra of various transmission components as functions of $\alpha$ with a fixed $\beta = 0.28$. At the resonant wavelengths corresponding to QBIC#1, QBIC#2, and the GMR mode in the $TM_1$, $TM_2$, and TE bands, respectively, the co-polarized components $T_{\mathrm{RR}}$ and $T_{\mathrm{LL}}$ exhibit significant differences. In contrast, the cross-polarized components $T_{\mathrm{RL}}$ and $T_{\mathrm{LR}}$ remain identical. This indicates that the CD originates from the difference in co-polarized components, suggesting that the chirality of this structure is intrinsic[37,38]. Furthermore, as $\alpha$ increases, the resonant positions of the GMR and QBIC#1 modes gradually approach each other. A distinct anticrossing feature emerges near $\alpha = 0.75$, indicating an interaction between them. Although the QBIC#2 mode also possesses chirality, its CD response remains weak ($|\mathrm{CD}|<0.7$). As $\alpha$ increases, the intrinsic CD at the resonant position of QBIC#1 is significantly enhanced in the vicinity of the anticrossing region. Near-perfect CD is achieved at 1653.9 nm when $\alpha = 0.91$. A sign reversal of the CD occurs at the resonant position of the GMR when $\alpha = 0.81$. Notably, the evolutionary trends of the QBIC#1 and GMR resonances, along with the significant differences in their CD responses, suggest a strong coupling between them. Detailed analysis of the definitive evidence for this coupling and its evolutionary laws will be presented in Section 3.2.

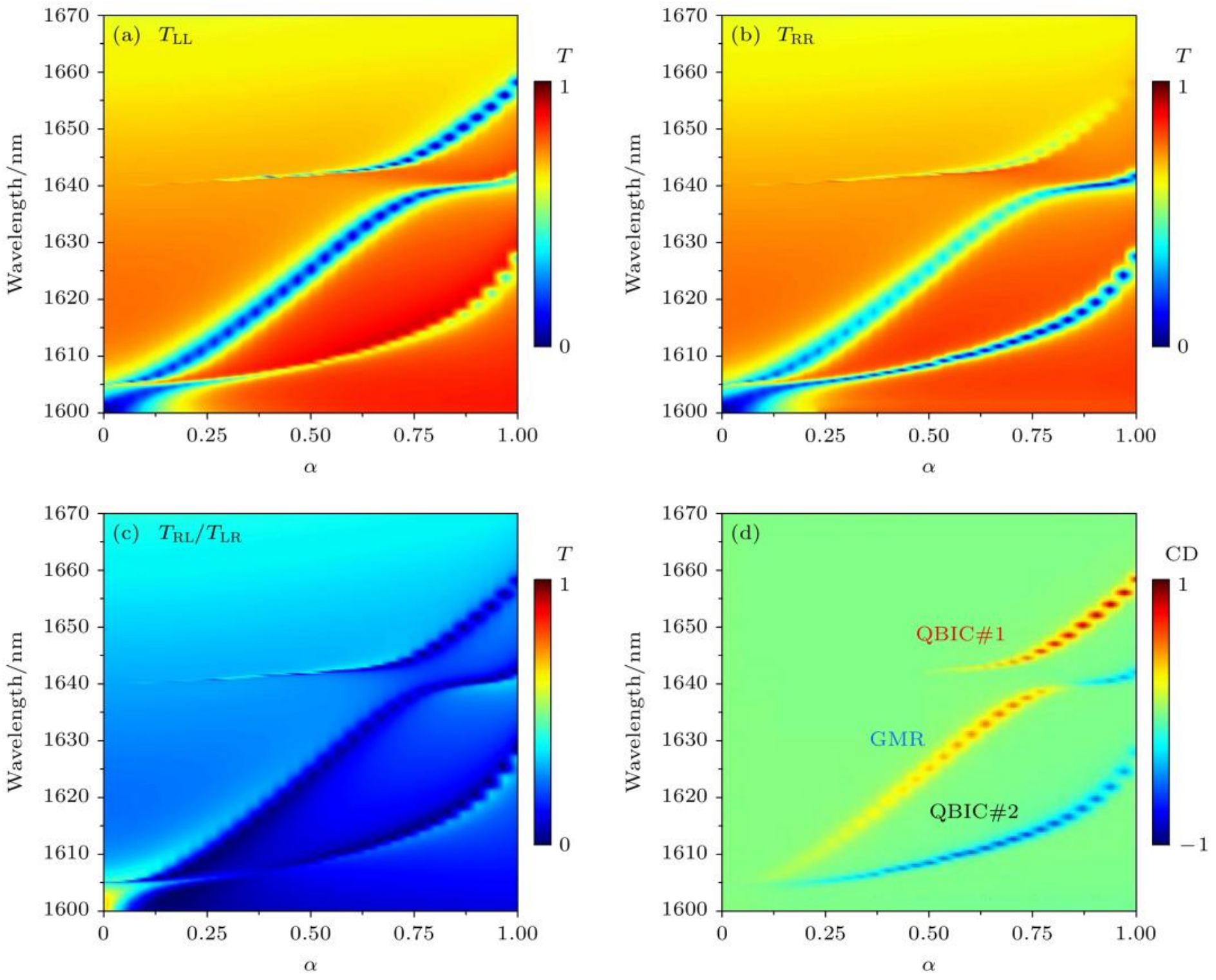

**Fig. 3 (a) Co-polarized $T_{LL}$ of transmitted LCP light and (b) co-polarized $T_{RR}$ of transmitted RCP light as functions of $\alpha$ with $\beta$ = 0.28; (c) cross-polarized $T_{RL}$ of transmitted LCP light and cross-polarized $T_{LR}$ of transmitted RCP light as functions of $\alpha$ with $\beta$ = 0.28; (d) CD responses as functions of $\alpha$ with $\beta$ = 0.28.**

## 3.2 Mode Coupling Characteristics

To elucidate the modulation mechanism of strong coupling on the chiral response of metasurfaces, we systematically investigated the coupling characteristics between the quasi-TM QBIC#1 mode and the quasi-TE GMR mode. In periodic metasurfaces, TE and TM modes possess orthogonal polarization eigenproperties. When their resonance frequencies converge closely ($\omega_{TE} \approx \omega_{TM}$) and their radiation losses are approximately matched ($\gamma_{TE} \approx \gamma_{TM}$), modal field overlap induces coupling. This condition is a prerequisite for achieving optimal energy exchange and interference effects between modes[39]. Figure 4 illustrates the evolution of the GMR (TE) and QBIC#1 ($TM_1$) modes as a function of the parameter $\alpha$, with $\beta$ fixed at 0.28. Significant coupling occurs near $\alpha$ = 0.75. In Figure 4(a), the resonant wavelengths exhibit typical anti-crossing behavior as $\alpha$ varies, with the resonant wavelengths of the two modes converging at this point ($\lambda_{GMR} \approx \lambda_{QBIC\#1}$). In Figure 4(b), the imaginary parts of the modal eigenfrequencies (corresponding to radiation loss $\gamma$) cross, indicating that $\gamma_{GMR} = \gamma_{QBIC\#1}$. These two features serve as key indicators of strong mode coupling[32]. Furthermore, Figure 4(c) displays the evolution of circular dichroism (CD) for the QBIC#1 and GMR modes during the coupling process. As $\alpha$ increases, the CD responses of the two modes change distinctly. In the region far from strong coupling ($\alpha < 0.5$), the CD value of the QBIC#1 mode is extremely low, approaching zero, whereas the GMR mode exhibits a positive CD. As $\alpha$ increases to approximately 0.75, the modes enter the strong coupling regime, and the CD undergoes dramatic evolution. Near $\alpha = 0.81$, the CD of the GMR mode reverses sign, shifting from positive to negative. In contrast, the CD of the QBIC#1 mode continues to strengthen, ultimately reaching a near-perfect CD of 0.995 at $\alpha = 0.91$. This evolution demonstrates that strong coupling not only significantly enhances the chiral response of the modes but also induces the inversion and reconstruction of chiral modes. Meanwhile, the $Q$-factors of the two modes in Figure 4(d) exhibit distinct evolutionary trends with respect to $\alpha$. The QBIC#1 mode, originating from a symmetry-protected bound state in the continuum (BIC), shows a gradual decrease in its $Q$-factor as $\alpha$ increases. At $\alpha = 0.91$, the mode maintains a high $Q$-factor of 1083 while achieving the aforementioned CD value. Conversely, the $Q$-factor of the GMR mode remains finite throughout. Notably, at the strong coupling characteristic point of $\alpha = 0.75$, the $Q$-factors of the two modes are equal, which corresponds precisely to the matching of modal radiation losses shown in Figure 4(b).

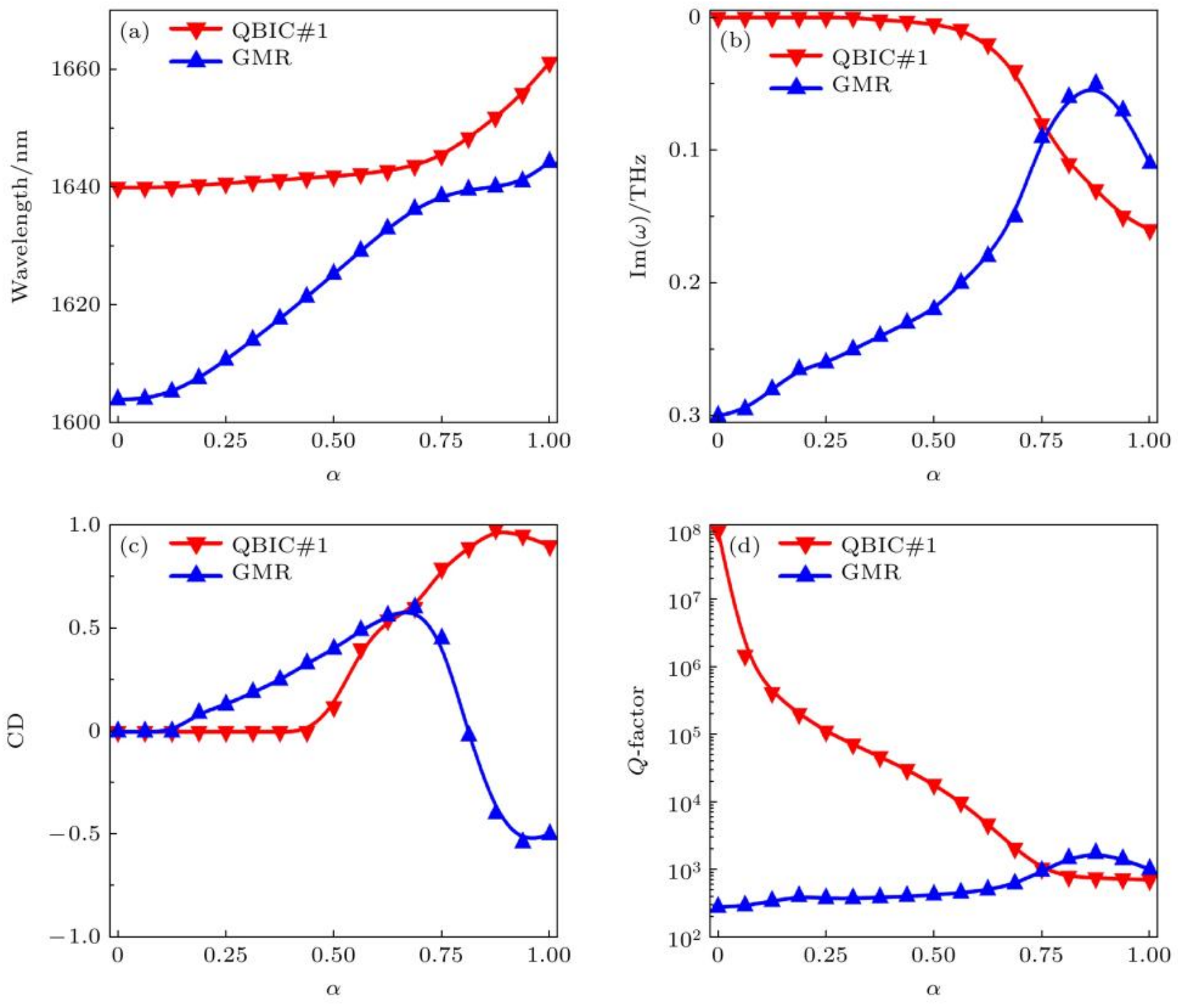

**Fig. 4 Variation of (a) resonant wavelength, (b) imaginary part of eigenfrequency, (c) CD, and (d) $Q$-factor with $\alpha$ for QBIC#1 and GMR modes with $\beta = 0.28$.**

Rabi splitting, defined as the frequency difference between high- and low-frequency hybridized modes under zero-detuning conditions, is a hallmark of strongly coupled systems. It arises from energy exchange and coherent interference between two resonant modes, with spectral anticrossing and energy splitting as its core features[40,41].

Therefore, for the coupled system composed of QBIC#1 and GMR, the energy eigenvalues can be obtained by solving the coupled-state equations[42]. Considering the energy exchange and loss between modes, the eigenvalue equation is expressed as

$$\begin{vmatrix} E_{\mathrm{QBIC\#1}} + \mathrm{i}\Gamma_{\mathrm{QBIC\#1}} & g \\ g & E_{\mathrm{GMR}} + \mathrm{i}\Gamma_{\mathrm{GMR}} \end{vmatrix} \begin{vmatrix} \alpha_{\mathrm{QBIC\#1}} \\ \alpha_{\mathrm{GMR}} \end{vmatrix} = E \begin{vmatrix} \alpha_{\mathrm{QBIC\#1}} \\ \alpha_{\mathrm{GMR}} \end{vmatrix}, \tag{2}$$

Here, $E_{\mathrm{QBIC\#1}}$ and $E_{\mathrm{GMR}}$ denote the resonance energies of the QBIC#1 and GMR modes, respectively, in the uncoupled state; $\Gamma_{\mathrm{QBIC\#1}}$ and $\Gamma_{\mathrm{GMR}}$ represent the radiative loss energies of the corresponding modes; $g$ indicates the coupling strength between the two modes. The eigenvectors of the system are expressed by the coefficients $\alpha_{\mathrm{QBIC\#1}}$ and $\alpha_{\mathrm{GMR}}$, which determine the hybridization weights of the two modes and satisfy the normalization condition $|\alpha_{\mathrm{QBIC\#1}}|^2 + |\alpha_{\mathrm{GMR}}|^2 = 1$; $E$ represents the eigenenergy of the hybridized modes, calculated as $E = \hbar\Omega$, where $\hbar$ is the reduced Planck constant and $\Omega$ denotes the resonance frequency. Solving Eq. (2) yields the energies of the upper and lower hybridized branches after coupling:

$$E_{\pm} = \frac{E_{\mathrm{GMR}} + E_{\mathrm{QBIC\#1}}}{2} - \mathrm{i}\frac{\Gamma_{\mathrm{GMR}} + \Gamma_{\mathrm{QBIC\#1}}}{2} \pm \sqrt{\left(\frac{E_{\mathrm{GMR}} - E_{\mathrm{QBIC\#1}} + \mathrm{i}(\Gamma_{\mathrm{GMR}} - \Gamma_{\mathrm{QBIC\#1}})}{2}\right)^2 + g^2}. \tag{3}$$

The Rabi splitting energy is defined as the difference in resonance energies between the upper and lower hybridized modes. It serves as a core physical quantity characterizing strong coupling strength, and its expression is given by

$$\Delta E = \hbar\Omega_{\mathrm{R}} = E_{+} - E_{-} = \sqrt{4g^2 + (E_{\mathrm{GMR}} - E_{\mathrm{QBIC\#1}})^2 + \mathrm{i}(\Gamma_{\mathrm{GMR}} - \Gamma_{\mathrm{QBIC\#1}})^2}, \tag{4}$$

Here, $\Omega_{\mathrm{R}}$ denotes the Rabi frequency. When the two modes satisfy resonant energy degeneracy ($E_{\mathrm{QBIC\#1}} = E_{\mathrm{GMR}}$) and matched radiative loss energies ($\Gamma_{\mathrm{QBIC\#1}} = \Gamma_{\mathrm{GMR}}$), the formula simplifies to $\Delta E = 2g$. In this case, the splitting energy is determined solely by the coupling strength $g$, corresponding to the optimal strong coupling state.

Figure 5(a) presents a stacked plot of CD spectra for different $\alpha$ values with fixed $\beta$ = 0.28. As shown in Figure 5, the CD responses of the two modes at the resonance positions of QBIC#1 and GMR exhibit significant differences as $\alpha$ varies. At $\alpha$ = 0.63, the CD signals of both modes are weak and show no sign difference. At $\alpha$ = 0.75, the CD value of QBIC#1 increases significantly to above 0.8. At $\alpha$ = 0.81, the CD response of the GMR mode begins to change sign. At $\alpha = 0.91$, QBIC#1 achieves a near-perfect CD of 0.995 at 1653.9 nm, while the CD of the GMR mode completely reverses to the negative range. This phenomenon intuitively reflects the differential modulation of chiral responses for different modes by strong coupling. Figure 5(b) shows the two-dimensional CD spectrum and Rabi splitting analysis results. The two-dimensional spectrum clearly displays a distinct anticrossing of the resonance wavelengths for the QBIC#1 and GMR modes near $\alpha = 0.75$, which aligns perfectly with the evolution curve of Rabi splitting characteristics. The consistency between the Rabi splitting evolution law and the quantitative correlation with CD response indicates that strong coupling between the QBIC#1 and GMR modes is the key mechanism for achieving high-$Q$ and high-efficiency intrinsic CD. As shown in Figure 5(b), at $\alpha = 0.75$, the central wavelength of strong coupling is located at 1642.5 nm, corresponding to a Rabi splitting energy valley of 3.16 meV. When $\alpha$ deviates from 0.75, the mode detuning $|E_{\mathrm{GMR}} - E_{\mathrm{QBIC\#1}}|$ and $|\Gamma_{\mathrm{GMR}} - \Gamma_{\mathrm{QBIC\#1}}|$ increase, causing a slight rise in splitting energy. This observation is consistent with the physical laws derived from the Rabi splitting energy formula. Calculation results indicate that the maximum Rabi splitting energy of this system approaches 7 meV under different $\alpha$ values.

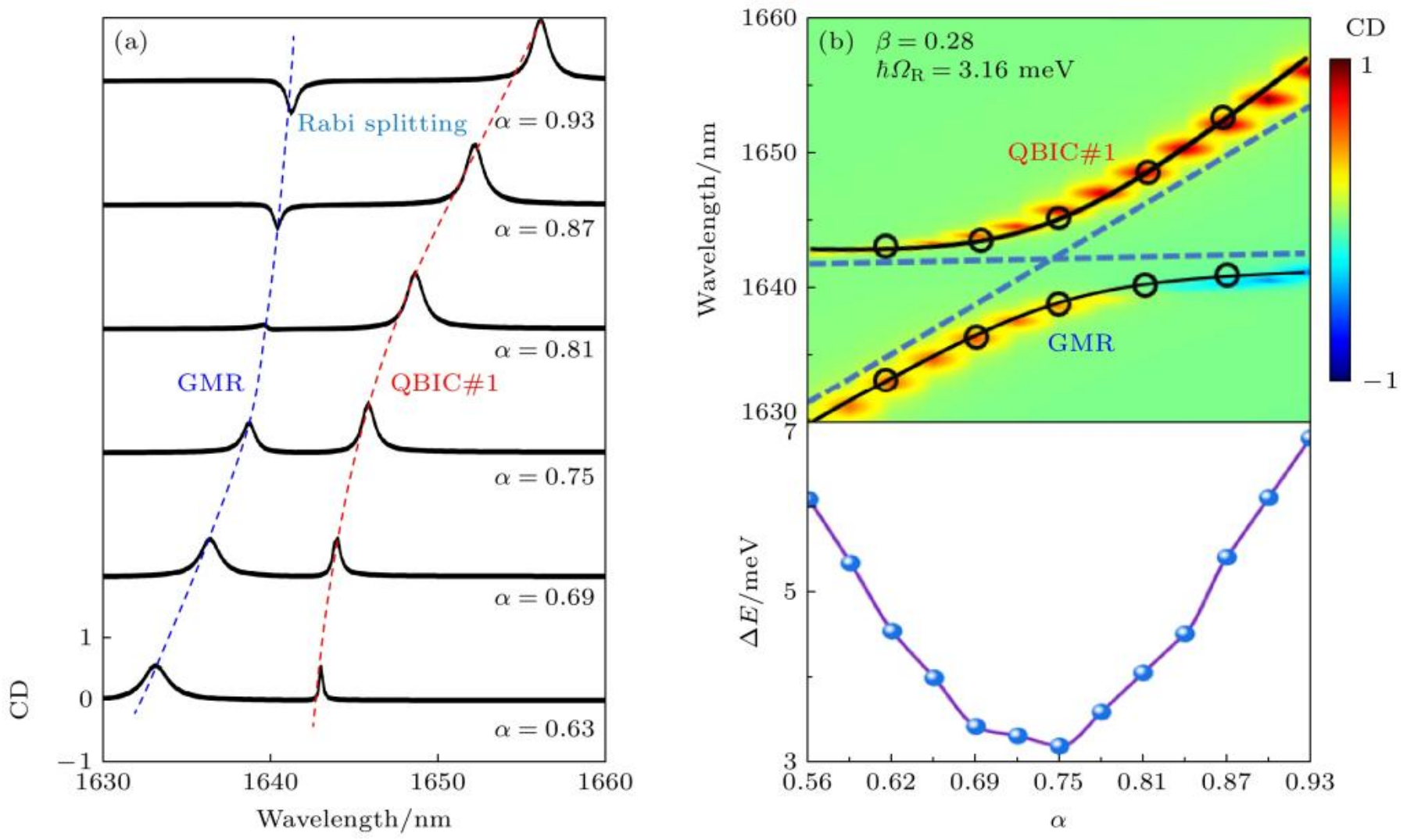

**Fig. 5 Under $\beta = 0.28$: (a) Stacked CD spectra at different $\alpha$ values; (b) CD spectra and Rabi splitting fitting results (top) and variation of Rabi splitting energy with $\alpha$ (bottom).**

### 3.3 Topological Evolution of Polarization Singularities

To elucidate the topological evolution of metasurface polarization singularities in momentum space under strong coupling mechanisms, we introduce the far-field polarization vector in the $x$-$y$ plane, $\boldsymbol{d}(\boldsymbol{k}_{//}) = d_x(\boldsymbol{k}_{//})\hat{x} + d_y(\boldsymbol{k}_{//})\hat{y}$. The topological charge $q$ carried by vortex polarization singularities (V points) or C points is defined as[43,44]

$$q = \frac{1}{2\pi}\oint_L \mathrm{d}\boldsymbol{k}_{//} \cdot \nabla_{\boldsymbol{k}_{//}} \phi(\boldsymbol{k}_{//}), \quad (5)$$

Here, $L$ denotes a counterclockwise closed loop in $\boldsymbol{k}_{//}$ space surrounding the BIC; $\phi(\boldsymbol{k}_{//})$ represents the angle of the polarization vector, where $\phi(\boldsymbol{k}_{//}) = \frac{1}{2}\arg[S_1(\boldsymbol{k}_{//}) + \mathrm{i}S_2(\boldsymbol{k}_{//})]$, and $S_1(\boldsymbol{k}_{//})$ is the Stokes parameter of $\boldsymbol{d}(\boldsymbol{k}_{//})$.

Figures 6(a)—(c) illustrate the evolution of the *C* point in the $TM_1$ band within momentum space as parameters *α* and *β* are tuned. As shown in Figure 6(a), for the unperturbed metasurface where (*α, β*) = (0, 0), the BIC manifests as a *V* point at the *Γ* point with a topological charge $q = +1$. Upon introducing the in-plane perturbation *α*, this *V* point splits into a pair of right-handed (RH) and left-handed (LH) *C* points, each with a topological charge of +1/2, thereby satisfying the law of conservation of total topological charge, as depicted in Figure 6(b). Furthermore, the out-of-plane perturbation *β*, introduced via under-etching, provides additional degrees of freedom for tuning this pair of *C* points. By simultaneously adjusting *α* and *β*, their positions in momentum space can be precisely controlled. Notably, as shown in Figure 6(c), when (*α, β*) = (0.91, 0.28), the LH *C* point is accurately tuned to the *Γ* point, marking the realization of an intrinsic chiral QBIC. The introduction of strong coupling fundamentally alters the evolutionary trajectory of polarization topology. When QBIC#1 and the GMR mode satisfy the strong coupling condition, their modal fields undergo deep superposition, reconstructing the spatial distribution of polarization vectors. The energy degeneracy induced by Rabi splitting drives the LH *C* point to gradually migrate to the center of momentum space, the *Γ* point. To visually demonstrate the changes in polarization states, Figures 6(d)—(f) further present the ellipticity angle *χ* distributions of the metasurface under different (*α, β*) conditions. The ellipticity angle *χ* is defined as $\chi = \arctan(n/m)$, where *m* and *n* represent the lengths of the semi-major and semi-minor axes of the polarization ellipse, respectively[45]. The value of *χ* ranges from −45° to +45°: $\chi = 0°$ corresponds to linear polarization, $\chi = +45°$ (−45°) corresponds to RH (LH) circular polarization, and other *χ* values represent RH or LH elliptical polarization. As shown in Figure 6(d), at (*α, β*) = (0, 0), the far-field radiation exhibits a linear polarization state with $\chi = 0°$ in momentum space. After introducing in-plane asymmetry ($\alpha = 0.91$), this polarization state is perturbed, generating RH and LH circular and elliptical polarization states, as shown in Figure 6(e). Finally, Figure 6(f) shows that upon further introducing out-of-plane asymmetry ($\beta = 0.28$), an LH circular polarization state ($\chi = -45°$) forms at the *Γ* point. These results indicate that strong coupling enables the precise tuning of the LH *C* point to the *Γ* point, thereby achieving intrinsic CD with both a high *Q*-factor and high efficiency.

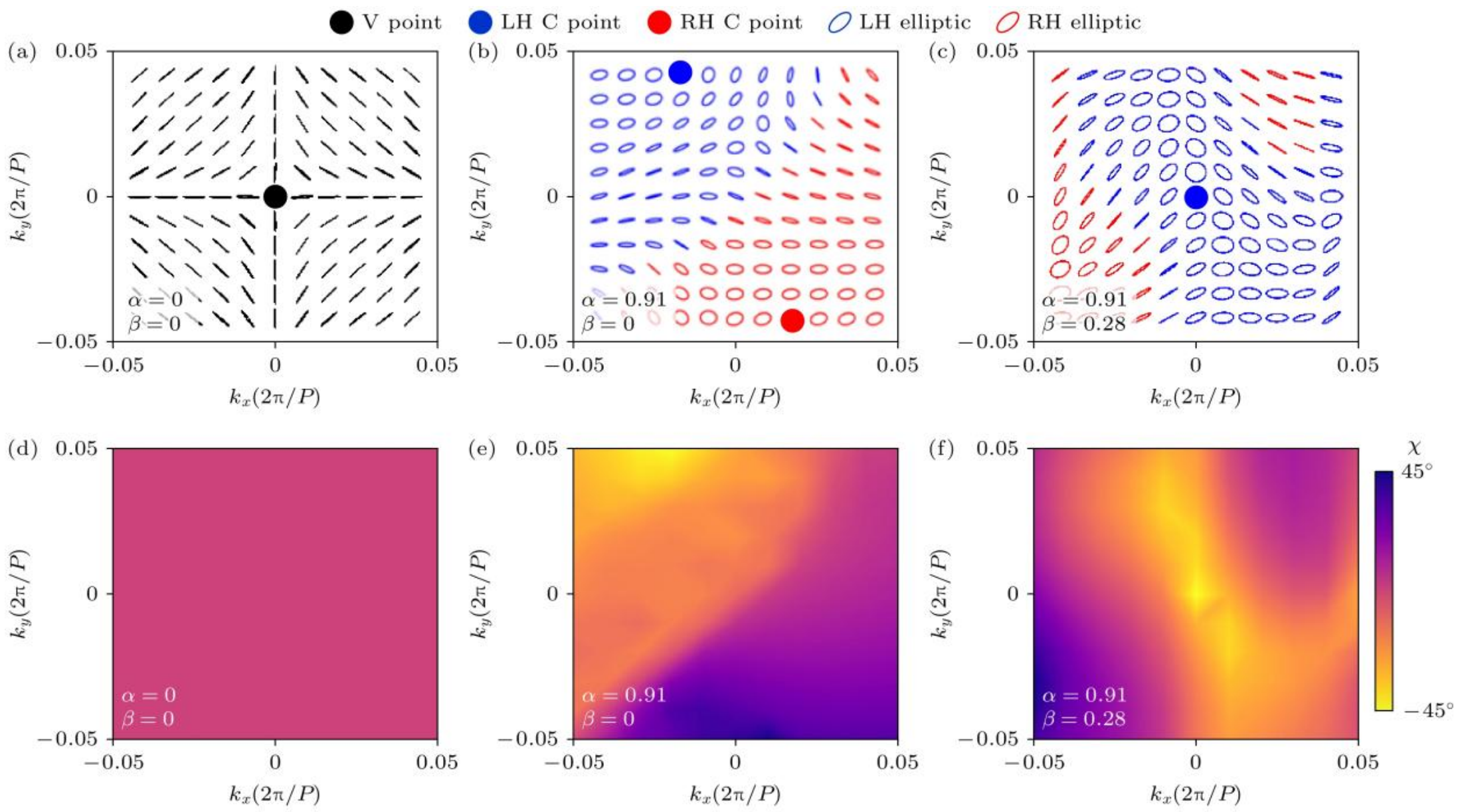

**Fig. 6 Distribution maps of *C* points of the $TM_1$ band in momentum space for (a) ($\alpha, \beta$) = (0, 0), (b) ($\alpha, \beta$) = (0.91, 0) and (c) ($\alpha, \beta$) = (0.91, 0.28); distribution maps of the ellipticity angle $\chi$ of the metasurface for (d) ($\alpha, \beta$) = (0, 0), (e) ($\alpha, \beta$) = (0.91, 0) and (f) ($\alpha, \beta$) = (0.91, 0.28).**

## 3.4 Influence of Structural Parameters on CD Response

Figure 7 illustrates the impact of variations in core structural parameters, including the unit cell period *P*, silicon layer thickness *h*, under-etching depth *d*, and groove width *w*, on the transmission CD spectrum of the metasurface. The results indicate that the metasurface maintains a stable high-*Q* and high-efficiency CD response even when these parameters vary over a wide range. This demonstrates that the intrinsic chiral states based on the strong coupling mechanism exhibit robust parameter tolerance. Furthermore, Figures 7(a)–(c) reveal that the chiral QBIC resonance peak undergoes a significant redshift as *P, h*, and *d* increase. This phenomenon occurs because the increased structural dimensions raise the overall effective refractive index of the metasurface, thereby lowering the resonance frequency. Conversely, an increase in *w* causes the resonance peak to shift toward the blue, as shown in Figure 7(d). This trend is primarily attributed to the reduced proportion of high-refractive-index silicon dielectric within the unit cell, which decreases the overall effective refractive index. Moreover, throughout the entire parameter variation process, the CD peak value remains at a high level (CD > 0.92), and the resonance linewidth does not exhibit significant broadening ($Q > 1000$). These findings suggest that the performance of this chiral metasurface is highly adaptable to common dimensional fluctuations during fabrication, providing a crucial foundation for its practical manufacturing and engineering applications.

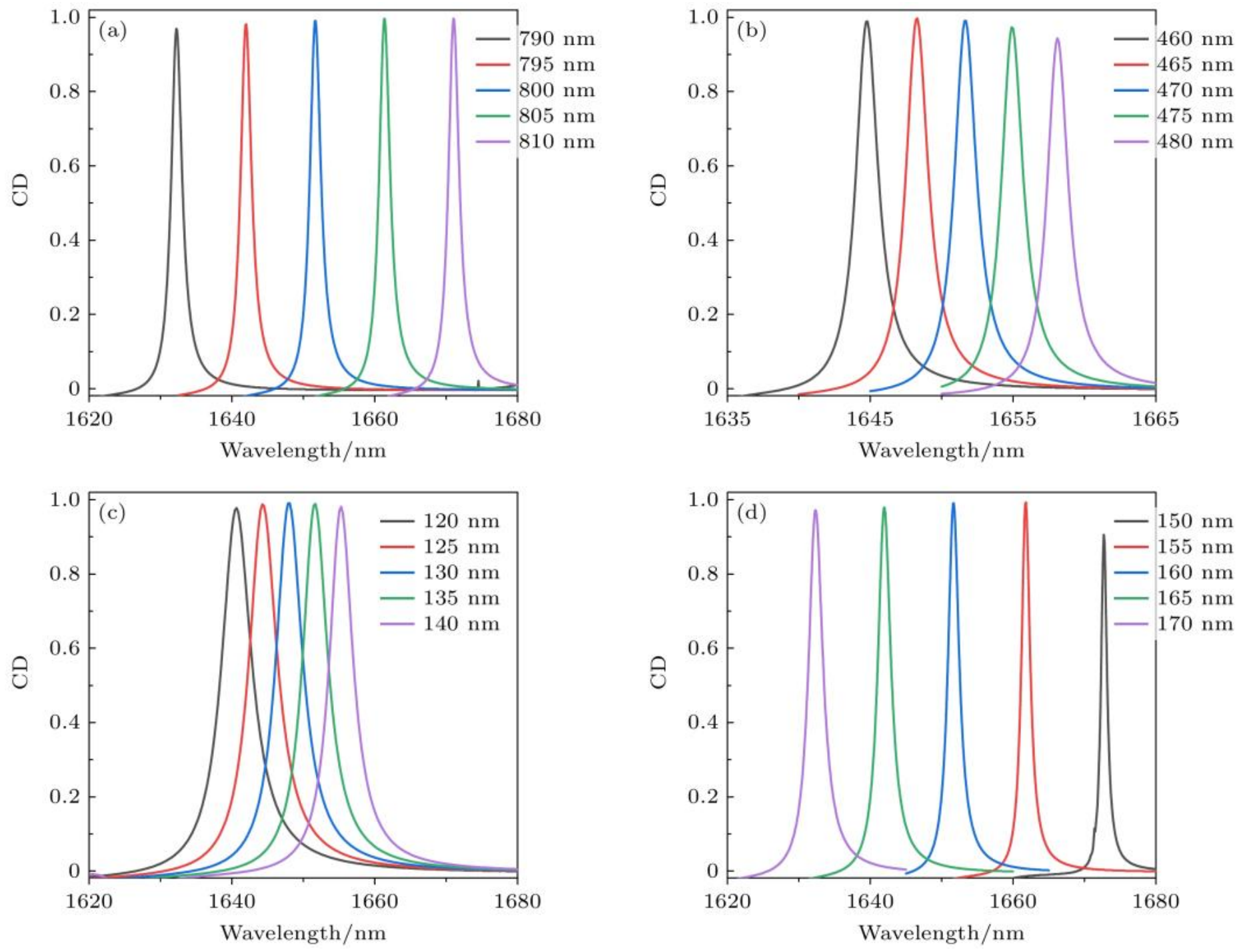

**Fig. 7 Influences of structural parameters of (a) *P*, (b) *h*, (c) *d* and (d) *w* on CD responses.**

# 4 Conclusion

This study constructs an intrinsic CD enhancement system based on the mode strong coupling effect in BIC metasurfaces. We designed a silicon-based cross-shaped nanopore array with $C_4$ symmetry. This structure simultaneously supports TM-like BIC modes and TE-like GMR modes within the communication band. Undercutting breaks the out-of-plane symmetry of the structure. Combined with systematic tuning of in-plane asymmetry parameters, this approach successfully induces strong coupling between the two modes. Simulation results confirm that near the strong coupling region, the QBIC#1 and GMR modes exhibit characteristic features of resonance wavelength anticrossing and crossing of the imaginary parts of eigenfrequencies. The corresponding Rabi splitting energy valley is 3.16 meV. The QBIC#1 mode achieves near-perfect intrinsic CD (CD = 0.995) with a $Q$-factor as high as 1083 at 1653.9 nm. In contrast, the GMR mode undergoes a significant CD sign reversal in the strong coupling region. This verifies the differential control of mode chirality by the strong coupling mechanism. Topological evolution of polarization singularities indicates that the strong coupling effect drives the LH $C$ point to precisely locate at the $\Gamma$ point in momentum space. This forms a left-handed circularly polarized state with $\chi$ = -45°. Robustness analysis of structural parameters shows that the metasurface adapts well to significant variations in unit cell period $P$, silicon layer thickness $h$, undercut depth $d$, and groove width $w$. The CD peak remains above 0.92, and the $Q$-factor stays above 1000. These results demonstrate excellent compatibility with fabrication processes. This study not only deepens the understanding of mode coupling mechanisms and chirality control in BIC metasurfaces but also provides feasible design concepts and theoretical support for applications such as high-precision chiral optical detection and high-performance chiral light sources.